# Sorry: Ambient Tactical Deception Via Malware-Based Social Engineering

**Adam Trowbridge, Jessica Westbrook, Filipo Sharevski**

DePaul University

## Biography

Adam Trowbridge is a designer, programmer, and code media researcher. Jessica Westbrook uses design to negotiate and organize the joys and struggles of information and understanding. Filipo Sharevski, Ph.D., is a cellular networks engineer and cybersecurity researcher. Together they co-direct Divergent Design Lab at DePaul University, focused on divergent thinking in emerging media practices.

## Abstract

In this paper we argue, drawing from the perspectives of cybersecurity and social psychology, that Internet-based manipulation of an individual or group reality using ambient tactical deception is possible using only software and changing words in a web browser. We call this attack Ambient Tactical Deception (ATD). Ambient, in artificial intelligence, describes software that is "unobtrusive," and completely integrated into a user's life. Tactical deception is an information warfare term for the use of deception on an opposing force. We suggest that an ATD attack could change the sentiment of text in a web browser. This could alter the victim's perception of reality by providing disinformation. Within the limit of online communication, even a pause in replying to a text can affect how people perceive each other. The outcomes of an ATD attack could include alienation, upsetting a victim, and influencing their feelings about an election, a spouse, or a corporation.

**Definitions**

As this paper relies on terminology from cybersecurity, but is presented outside that context, specific terms should be available to the reader. *Cybersecurity* is "[t]he approach and actions associated with security risk management processes followed by organizations and states to protect confidentiality, integrity and availability of data and assets used in cyber space" (Schatz, Wall, and Wall 2017). The term *social engineering* refers to the use of social interaction to gain information, or manipulate a situation, as part of hacking or bypassing a system. This paper uses the term *malicious actors* to describe people who engage in what is generally called "hacking." In this paper, *attack* refers to an attempt to hack a system, or to socially engineer a person, and *victim* refers to a person or entity that is the target of an attack. A *daemon* is a computer program that runs as a background process, rather than under the direct control of an interactive user. The term *malware* refers to software that is intended to damage, alter, or disable computers and computer systems.

## Introduction

At one time, Susan Headley was a phone hacker ('phreak'). She claims to have been able to use communication, on the phone and in person, to obtain information about the work schedules of intercontinental ballistic missile sites. This sort of attack is part of "social engineering," in which a malicious actor attempts to introduce false information into a conversation in order to gain access to data or a network. Social engineering has often been a counterpart to "hacking" computers and networks. Her abilities as a hacker and phone phreak were surpassed by her ability to combine affect and communication into a tool for manipulating reality, in order to gain access to systems and information (Hafner and Markoff 1995). Headley and other early hackers discovered that merely saying they were calling from a specific company or organization seemed *true enough* that secretaries were willing turn over valuable information like names of people who work on specific projects. Further investigation, which might include diving into the dumpster for company communication and technology manuals, would reveal personal details about those people, hardware and software manuals, and phone lists.

While "exploits" in cybersecurity often involve mistakes in complex code, the tactics with which some malicious actors exploit systems are very human: social and political. All other forms of hacking involve specific input into a computer and specific output from the computer, a machine that cannot accept anything but specific input and deliver anything but specific output. Social engineering, as employed by Headley and other social engineers, is much more nuanced, but hacking people still requires an exploit, a weakness, or some sort of vulnerability.

Deception involves manipulating the reality of an individual: the social engineer introduces herself as a friend of another employee, and requests that employee's address in order to send a gift. The assistant who answers the phone is given false information based on seemingly reliable data, primarily because the person on the phone seems truthful. Having another person act falsely taxes the person being targeted for social engineering. It causes a dilemma: personal information should not be provided to strangers on the phone, but it would be rude not to provide an address to someone attempting to send a gift, particularly when the person on the phone seems distressed over forgetting. The target of a social engineering attack must resist, if they resist at all, from the basis of questioning what another person claims is reality. As seen below with examples of phishing and manipulating text, how likely someone is to resist social engineering is a complex mix of learned behavior and social norms.

The social engineer begins with a few bits of information: a name, phone numbers, perhaps dumpster-borne memos that reveal the company organization chart. She selects a target from the org chart, uses these bits of information to call an employee's assistant, and asks for a home address. The assistant is resistant to social engineering, and despite the possibility of seeming rude, refuses to provide the home address. Feigning forgetfulness at points in the conversation, the social engineer also casually asks the assistant the name of the target's dog, and the bar the target frequents. These bits of information seem much more benign to the assistant. After hanging up, the assistant remains unaware that the conversation was not intended to reveal an address (which is easily found on the Internet), but, instead, personal details about the pet and the bar (which were not available on the Internet). The assistant gave them out freely. Each new bit of information leads to more information on which to construct new social engineering realities, until the attacker has enough information to approach her primary target or attempt to hack into their accounts.

Humans invented computers in a coordinated effort that constructed a useful abstraction of reality. That reality is governed by code, sets of instructions and rules. The device on which I am typing these words, a MacBook Pro 2015, can do virtually anything. The computer in my pocket, a smart phone, has at least eighteen times more processing power than a 1985 Cray-2 supercomputer (Experts Exchange 2018). What my computer cannot do itself, it can do when networked to similar devices, some of which are networked to physical devices, environments, communications networks, broadcast networks, military installations, corporations, and government facilities. As I type on this device, I can communicate with virtual any willing party in the world, at any time. However, I may also choose to type things that allow me to access virtually everything, including secure systems to which I have not been granted access. There is a lure to testing out these networks, to obtaining information that is forbidden. I may also choose simply to write an email to a co-worker. That form of communication is made up of digital

"packets" that contain data on the use of the packet (i.e. where to send it) and the data to be sent. There is no difference between the text in a Word document and the text sent across the Internet. It all reduces to data, made up of numbers.

When the text in my email arrives, the data is moved from one set of numbers to another, and to another, moving from the packets of data into organized information. This may be displayed on a browser window via an online email program. At any point bad data may be introduced into the system, from the packets to the programs that translate that data into information, randomly or by intent. Some interruptions and diversions of data or information may be caused by a malicious actor. Some of these people, commonly known as "hackers," divert packets, primarily to obtain data. Other malicious actors construct their own bits of reality in the form of email messages that attempt to deceive people into providing access to even more packets of data, such as those containing credit card numbers.

Internet packets may also be intercepted between one computer and the next, changed, then released, introducing doubt into the system. People tend to trust their email, but any email could be changed on the computer of the sender, *en route* across the Internet, or on the computer of the receiver. An email could be from a malicious actor, pretending to be a friend or a bank. Fake emails can be designed for the purpose of deception, and regularly are. However, the content of an authentic email can also be changed, and possibly without the notice of the reader. Words can be deleted and added, changed slightly or completely, and those changes can shift what little affect is available in an email message. Like the assistant's dilemma above, how people read email is influenced by social norms.

This paper will propose a potential malware attack that could purposefully change, remove, and add words to everything that appears on a target's browser, and potentially all the text on a computer or device. This attack would combine the vulnerability of text stored in computer data with people and social systems, and could be buttressed by automation via artificial intelligence. Due to contemporary reliance on Internet communication, and the ubiquity of web browsers as interfaces for using the Internet, many people would be vulnerable to this type attack. This paper discusses which communications are vulnerable, how they are vulnerable, and how they could be exploited to manipulate a target's individual and social reality. Relevant existing examples of reality-interrupting software are presented, as well as a prototype system developed by the authors of this paper.

## Descartes's Demon

Philosopher Rene Descartes imagined a demon who was able to shape his reality:

> "*I will suppose therefore that...some malicious demon of the utmost power and cunning has employed all his energies in order to deceive me. I shall think that the sky, the air, the earth, colors, shapes, sounds and all external things are merely the delusions of dreams which he has devised to ensnare my judgement.*" (Descartes, Ariew, and Cress 2006)

Descartes's demon can create entire realities, leaving Descartes sure of nothing but his own existence. This demon has absolute power over Descartes's senses, everything known could be an illusion. However, what if the demon was much less powerful, but still had access to people's reality? This *lesser* demon can only apply a limited amount of power, and only to a small number of people. This demon controls what a victim hears when other people are speaking, and can also change the victim's responses as heard by other people. The power is fleeting, and if the victim grows too suspicious, the illusion dissipates, leaving the victim aware of unaltered reality.

The lesser demon can change the affect of what is said by swapping out specific words, or change a sentence completely. If he goes too far, any party in a conversation may grow suspicious. Friends may show concern that what the victim is saying makes no sense. The lesser demon must ensure that what the victim hears flows with what the victim expects to hear, and that what the victim says flows with the conversation. The demon must maintain the flow of social reality so that what everyone hears is within the scope of what they expect to hear.

The only advantage of such a power would be to subtly influence a victim, to prod the victim generally toward actions, or away from specific people. It would not be possible to make a victim do what the demon wanted, directly, but over time small changes could add up to a victim with a significantly different (and incorrect) perception of social reality. The demon could lead a victim to distrust a specific person, and to take what that person said in the wrong way. This could end friendships, and possibly change a victim's thinking, their outlook, their politics, and their close relationships.

Due to the massive amount of electronic communication occurring over networks, all of which are vulnerable to hacking, malicious actors could act in ways similar to a lesser demon. Their ability to influence a victim would be undone if they attempted to create drastic changes to reality. They could, however, make small changes to text, even if only within web browsers, and still manage to manipulate the victim's perception of reality. The technology to do this currently exists, and could be applied to influencing a person, a group, or an election.

**Jeremy Corbyn in a Vat**

Billions of dollars of venture capital are currently invested in the belief that it is possible to develop communication delivery channels to reach very specific groups of people, and to tailor advertising messages so that they appeal to specific people. This "micro-targeting" is the capital-oriented, driving force behind social media websites that users perceive as free. To what degree could this sort of customization shape an individual's reality?

According to a source in Tom Baldwin recent book *Ctrl Alt Delete: How Politics and the Media Crashed Our Democracy*, Labour Party campaign chiefs were asked by Jeremy Corbyn to run a series of ads (Baldwin 2018). They believed these ads were too expensive, and did not run them widely. Instead, they ran the ads so that only Corbyn and his team would see them, online, using "individually-targeted, hyper-specific ads made possible through Facebook's advertising tools." In essence, they manipulated Corbyn's perception of reality. A Labour party official described the manipulated reality: "If it was there for [Corbyn and his associates], they thought it must be there for everyone" (Haskins 2018). The story has since been denied by people in the Labour party, but the manipulation of reality described is not only technologically possible, delivering ads to very specific groups is exactly the goal of micro-targeted advertising.

The outlook of Corbyn and his staff, that what they see must be true for everyone, that what they perceive is correct, is best described, via Truth-Default Theory, as Truth-Bias, the tendency to actively believe or passively presume that another person's communication is honest independent of actual honesty; and Truth-default, a "passive presumption of honesty due to a failure to actively consider the possibility of deceit at all or as a fall back cognitive state after a failure to obtain sufficient affirmative evidence for deception" (Levine 2014). This theory describes human communication, but may also describe people's relationship to their tools. In a paper on deceptive emails, Williams and Polage suggest that "[w]ithout a reason to doubt the legitimacy of an email, participants may then simply defer to assuming that the communication is likely to be genuine" (Williams and Polage 2018). In many situations, some of which are discussed below, people assume that what they see on a screen and within a browser window has not been altered. They default to a belief that their tool is acting as it should, and it is providing information that is "true," insomuch as the text is the same as what was written by the author. Cybersecurity describes this accurate state of data as *integrity*: the "information is not altered, and that the source of the information is genuine." *Confidentiality*, or "protecting information from being accessed by unauthorized parties;" *integrity*; and *availability*, or that "information is accessible by authorized users;" form the "CIA" triad in cybersecurity, which seeks to protect all three facets of information transfer (Mozilla 2018).

### Social Engineering

Social engineering, as defined above, requires the use of social interaction to gain information, or manipulate a situation. Social engineering is a term specific to security and cybersecurity, but it describes activity that is also researched in criminology and social psychology. At its base, social engineering uses deception as part of social interaction to get information, for example: making a call to tech support pretending to be a specific user who has forgotten their password. Social engineering attacks could also be considered a "con," or, legally, fraud.

Phishing is a social engineering attack that most people who work in a large organization hear about regularly from their information technology staff, even if many remain unfamiliar with the term. In a phishing attack, a user receives an email from a known entity, for example: from their IT department, their bank, or a social media account. That email includes a call to action and a link. People click on the link and are met with a login screen asking for their user name and password for that account. The website on which they enter this information is not what it seems. It has only been set up to collect the usernames and passwords entered. Every year millions of people fall for phishing scams, and in doing so, provide malicious actors access to their account. Regarding the use of social engineering in email phishing, Opazo et al. write "Existing security systems are not widely implemented and cannot provide perfect protection against a technological threat that relies on social engineering for success" (Opazo, Whitteker, and Shing 2017). Most dramatically, the paper declares that if the phishing scam is convincing, users are "generally helpless."

Phishing is based on avoiding what Truth-Default Theory calls "Deception judgment." In Timothy R. Levine's *Truth-Default Theory (TDT): A Theory of Human Deception and Deception Detection*, he writes:

> "*If a trigger or set of triggers is sufficiently potent, a threshold is crossed, suspicion is generated, the truth-default is at least temporarily abandoned, the communication is scrutinized, and evidence is cognitively retrieved and/or sought to assess honesty-deceit*" (Levine 2014).

To avoid these triggers, malicious actors employ legitimacy-by-design: design that appears legitimate based on what people expect to see. Phishing emails use actual company logos, and are based on email templates that seem as if they came from a trusted source. They use language as close as possible to the language the victim expects to see from that source. In *Why Phishing Works* Rachna Dhamija et al. point out that research intended to foster online trust among consumers does not consider that the same approaches could be used by malicious actors. Their study showed that even when users were aware of the potential for phishing, they were unable to detect phishing emails. In their study, "the best phishing site was able to fool more than 90% of participants" (Dhamija, Tygar, and Hearst 2006).

**Man-in-the-Middle**

Above, Descartes's lesser demon can control what a victim hears when other people are speaking, can change words spoken, and can also change the victim's response as heard by another party. In cybersecurity this scenario would be classified as a *man-in-the-middle* attack. It is frequently employed by a malicious actor to obtain information. In communication over a network (including the Internet) the person in the middle may be able to capture, alter, and release communication. A simple, capture-only form of this attack in daily life is eavesdropping, when someone listens in on a conversation to which they are not a party. In cybersecurity, this approach can be used to electronically "listen" to a "conversation" between a user and a website server they are logging into, giving the attacker a username and password.

It is possible to act as a man-in-the-middle using malicious software, and, instead of merely "listening," craft a different reality for the victim by changing text in their web browser. We call this new form of attack, drawing from the perspectives of cybersecurity, information warfare, and social psychology, Ambient Tactical Deception (ATD). Specific words could be changed, removed, or added that change the tone of an email, or a web page. Ambient, in artificial intelligence, describes software that is "unobtrusive," and completely integrated in a user's life. Tactical deception is an information warfare term for using deception on an opposing force. These terms, combined, differentiate ATD from existing man-in-the-middle attacks. Instead of seeking to gain advantage over a victim by stealing login credentials, an attacker might seek to influence the victim's perception of reality by providing disinformation. The outcomes of an ATD attack could include alienation, upsetting a victim, and influencing their feelings about an election, a spouse, or a corporation.

**Ambient Tactical Deception: Ambience**

The term "ambient" has also been used in the ambient music genre. It describes music that occurs in the background, but also describes activity more concerned with atmosphere and mood than specific structure or rhythm. A successful ATD attack must not be noticeable by the victim, for at least some time period. Changes remain in the background, so the victim is unaware that text is being altered. Ambient is also used in artificial intelligence to describe AI that analyzes people, things, and their environment so that it can instantly make predictions and provide control (Sadri 2011). The same technology that analyzes and creates human affect, a field called "affective computing," could be used for an advanced ambient tactical deception attack. Within the limit of online communication, even a pause in replying to a text can affect how people perceive each other (The Conversation 2014). It is possible that automated software could capture and change not only text, but the tone or mood of the text.

In some experiments, social scientists need to induce moods in people taking part in research. This often includes the use of movies and images. In an increasingly online age, researchers have investigated whether text alone can alter a person's mood. Verheyen and Göritz researched this with specific texts and found that "the texts effect a genuine mood change." (Verheyen and Göritz 2009). Their results also supported the previous finding that "negative mood was induced more effectively than positive mood." In a different study, Göritz found that the use of "affectively valenced words" alone was not able to produce a positive or negative mood (Göritz 2007).

Gendron and Barrett describe emotions as "dynamic multidimensional events," and write that the "perceiver must extrapolate from subtle, variable, and dynamic movements and utterances, embedded within a situation, to arrive at an understanding of another's continually evolving internal state" (Gendron and Barrett 2018). Email communication is asynchronous, and contains little opportunity for perception of emotion outside what can be pulled from the text. Email text is often short. Group emails accumulate quickly. In a study of communication of workplace email, Kristin Byron notes that "the reduced availability of cues and feedback may make email communication in general less physiologically arousing than face-to-face interaction" (Byron 2008). She posited a "neutrality effect" and "negativity effect" of email communication, in which positive emotion is dampened by email, while negative emotion is intensified.

In "Linguistic Politeness in Student-Team Emails: Its Impact on Trust Between Leaders and Members," the authors examined email directives that are requests for the person who received the email to do something (Lam 2011). People in work environments often need to issue directives to the people who work for them. Issuing directives is a face-threatening act, which can "cause addressees to lose face by either imposing on an addressee's autonomy or imposing on an addressee's desire to be included, appreciated or liked." For the directive "complete the budget report," used as an example in the paper, they found evidence that two variations "mitigate the force of the speech act:" (1) "I understand that you are extremely busy these days, but can you complete the budget report?" and (2) "Would it be ok for you to possibly complete the budget report?" Example 1 is referred to as a supportive move, it recognizes that the person receiving the email is busy, before issuing the directive. Example 2 is referred to as a "downgrader," it softens the directive by adding "would it be okay…?" to form a question. These approaches were found to support building trust. A third modification of the directive, "Unless you want to lose points, can you complete the budget report?" is an aggravating move. It makes the directive more apparent, adds a direct threat, and was found less effective at building trust. The authors write, "Across all theories of linguistic politeness, the basic foundation of each theory is the notion that a speaker's language choices have the power to impact interpersonal relationships (Lam 2011).

### Ambient Tactical Deception: Words

An email ATD attack relies on words: adding, removing, or changing them. As discussed above, this attack could focus on any text on a computer or computer network. The remaining discussion of ATD in this paper will focus on email communication via an online email reader (e.g. Gmail and Outlook on the web) and one particular situation: changing email communication on one person's computer to change how they feel about that person's communications, to some degree. As discussed below, this is a particularly vulnerable form of communication, and perhaps an attractive target. In order to maintain the "ambience" of the attack, the malicious actor will need to change text as little text as possible, for maximum effect. For that, they can pull from academic work from multiple fields on positive and negative words.

In developing their list of valenced words, Danion et al. used a dictionary, selecting words that 100% of their research agreed were valenced as positive, negative, or neutral. The valence of these words was then rated by undergraduate students (Danion et al. 1995). This approach could be combined with the approach in "Linguistic Politeness in Student-Team Emails: Its Impact on Trust Between Leaders and Members," which used a pragmatic politeness taxonomy (Lam 2011). There is ample research that could be combined to develop a "starter set" of words which would not stand out in email communication, but that could change the affect of the content.

Mackiewicz and Riley state, in *The Technical Editor as Diplomat: Linguistic Strategies for Balancing Clarity and Politeness*, "pragmatics is the branch of linguistics concerned with how language use and interpretation are affected by specific contexts." This area of research overlaps much of the communication focus of ATD. The paper suggests that pragmatics "can help editors communicate more effectively by using specific linguistic strategies to balance clarity and politeness (Mackiewicz and Riley 2003). In the case of ATD email attack, a malicious actor seeks to effectively employ linguistic strategies to shift the interpersonal reality of a target. In order to alienate a target from someone, the tone of that person's emails could be edited to change or remove words with positive affect, and include words with negative affect. As discussed above, some research supports the idea that adding negative affect to an email might create a negative feeling in the target that is not in proportion to the feeling that same affect would create in person-to-person communication.

### Ambient Tactical Deception: Tactical Deception

Tactical deception is an information warfare term for using deception on an opposing force. In *Shannon, Hypergames and Information Warfare*, Carlo Kopp defines "Four canonical offensive Information Warfare strategies": Denial of Information; Deception and Mimicry; Disruption and Destruction; and Subversion. Of these, deception and mimicry, or "mimicking a known signal so well, that a receiver cannot distinguish the phony signal from the real signal" is key to ATD, and has already been discussed in our use of the word "ambient" (Kopp 2003). Deception and mimicry are also important in response time: the ATD attack cannot cause the target's computer or software to act significantly differently, including the speed it takes to load a page. Disruption is "insertion of information which produces a dysfunction inside the opponent's system." In the case of an email ATD attack, the opponent's system is their perception of reality as it relates to the email author. The ATD attack could interrupt or disrupt internal communications within any social system based primarily on electronic communication. Loosely affiliated social and political groups that primarily rely on Google Suite tools, social media, and email are particularly vulnerable, as are companies that rely primarily on off-site, online workers for vital functions.

As described by Lin and Kerr, Information/Influence Warfare and Manipulation (IIWAM) is "the deliberate use of information against an adversary to confuse, mislead, and perhaps to influence the choices and decisions that the adversary makes." They suggest that a "cyber-enabled" IIWAM could exploit "modern communications technologies to obtain benefits afforded by high connectivity, low latency, high degrees of anonymity, insensitivity to distance and national borders, democratized access to publishing capabilities, and inexpensive production and consumption of information content" (Lin and Kerr 2017). The paper is an excellent review of contemporary information warfare, but it neglects to account for the micro-targeting capabilities of contemporary networks and social media. Using only information available publicly, or adding social engineering to the attack preparation, a malicious actor could know a great deal about a victim's social network and which relationships to target. ATD would seem to be a subset of cyber-enabled IIWAM, but the tools described in the paper are blunt, and aimed at masses of people, whereas ATD would be aimed at individuals, or a small network of individuals.

### Ambient Tactical Deception: Vulnerabilities

Lin and Kerr identify the most likely targets of cyber IIWAM as "users that have abandoned traditional intermediaries," such as newspapers and other sources that include editorial judgement of the information provided. They describe these people as tending "to be exposed preferentially (or almost exclusively) to…information that conforms to their own individual preferences." They also note that people who rely on social media and search engines for news are "less likely to be exposed to information that contradicts their prior beliefs" (Lin and Kerr 2017). This template can be adapted to describe potential victims of an ATD attack. Any relationship in which people rely on web-browsers for email, communicate nearly-exclusively via email and short message service (SMS) on a smart device, and have little or no contact with those specific people outside electronic communication would be excellent targets for an ATD attack.

The nature and potential of an ATD-type attack in international cyber-warfare, or attempted political influence, warrants the attention of cognitive scientists and researchers. Hundreds of millions of people receive much of their communication and outside information via a web browser (Perrin and Jiang 2018; Dimmick, Chen, and Li 2004). Online news sources, browser-based email, social media, and deep-web services like university and corporate intranets shape much of people's view of reality on a daily basis. Research has shown that citizens of the US and UK prefer text messaging over talking (Informate Mobile Intelligence 2015). 67% of Americans get at least some of their news from social media (Shearer and Gottfried 2017), and some people work in situations where their only contact is via the Internet (Rosenberg 2017). These all describe a population of potential victims who may not have readily available outside confirmation as to whether what they read on the web is true.

### Ambient Tactical Deception: Technological Plausibility

Technically, there are two types of ATD attacks that can or have been waged online. The first is ad-based, where a sophisticated adversary uses background information of targeted victim's preferences to craft special ads on social platforms with the intention to influence (1) political opinion (Cambridge Analytica scandal) or (2) maintain an alternative reality perception (e.g. micro-targeting Jeremy Corbyn and his associates). In the Cambridge Analytica case, a political data firm gained access to private information on more than 87 million Facebook users (Kozlowska 2018). The firm then offered tools that could identify the personalities of voters and influence their political opinion.

The ad-based ATD attacks require access to private information and sensitive confidential documentation. Instead, an ATD attack based on malware requires only a small piece of software in a form of a browser extension. It is possible for malicious actors to develop such an extension (see next section) and deliver it to the targeted user via social engineering or directly installing on their devices. Both the ad-based and malware-based ATD attack can be also used for micro-targeting. The malware-based attack provides an advantage to inducing certain moods or creating an alternative reality (with prolonged usage) because it is hard to detect. For the malicious actor, there is no threat that the target might use ad-blockers or simply ignore targeted ads. The malware-based ATD attack, if deployed successfully, is independent of the source of the webpage or the social media platform the targeted user is using. A malicious actor only needs an extension for popular browsers that is capable of changing text. Such extensions exist; for example, there is a Chrome extension that replaces every mention of the words "Elon Musk" with "Grimes's Boyfriend." The result is that people with this extension installed are read headlines as "Is Grimes's Boyfriend just an AI set on 'eccentric billionaire' mode?" and "Grimes's Boyfriend plans to create bricks for affordable housing" (Vincent 2018).

The first step in an ATD attack is install the malicious browser extension on the targeted user's system first. That can be done by cloaking it as a standard utility, using names like "Stickies" and "Lite Bookmarks" (Newman 2018). This will work for two reasons: (1) Developing extensions for Chrome is free; A benign extension can be submitted for publishing and pass all the security checks at Chrome. If the extension is installed on the targeted user's system, it will ask permission to change text, to allow for copy/paste. In a study of Android users, that may point to overall computer user behavior, researchers found that only 17% of users paid attention to permission granting when installing apps (Felt et al. 2012). Later, it is trivial to change the behavior dynamically, and use these permissions to alter any text as part of the malware-based ATD attack. (2) Chrome is already a trusted application; When users give it permission to run certain code, like an extension, their operating system and most antivirus products usually give it a free pass.

## Ambient Tactical Deception: Precursors

There are categories of existing software that change web pages in ways analogous to an ATD attack. Facebook Purity (FB Purity) is a web browser extension that allows users to customize Facebook. While it has been discussed as allowing users to block "annoying" Facebook features (Gordon 2010), it currently includes a feature called "Text Filter," with the instructions "Enter the words or phrases, on separate lines, that you wish to filter from your news feed." FB Purity will then block any post or response that contains the words the user enters. This feature has been advertised as a way to block political posts from a user's Facebook feed (FB Purity 2012). In essence, FB Purity allows a user to customize their reality when they use Facebook, and to remove from that social media reality posts that discuss topics they would like to avoid. Facebook's own Ad Settings allows users to hide ad topics related to alcohol, parenting, and pets. This feature, added in 2016, was intended to help people avoid topics that might upset them (Sloane 2016).

While FB Purity attempts to improve Facebook in ways that some users appreciate, artist Ben Grosser's extensions intentionally disrupt the core user experience on Facebook. His *Facebook Demetricator* removes all metrics (e.g. number of likes, number of friends, number of comments) from Facebook (Grosser 2018). Grosser later released *Twitter Demetricator* which takes the same approach to Twitter's metrics. While these projects are a critique of the social media construct they modify, they also actively disrupt the social reality structure imposed by Facebook in their interface choices, and, to some degree, create a new experience of social interaction on those platforms.

The primary inspiration for the prototype ATD software discussed below is browser extension *Jailbreak the Patriarchy* (JtP), developed by Danielle Sucher. *JtP* swaps the gender in browser texts, based on a list of 299 gendered terms (Sucher 2015). King becomes queen, actress becomes actor, duchess becomes duke, and her becomes his. Sucher said that the project was born from a discussion about eBooks. She considered what possible advantage could come of having access to the text of the books on a computer and decided to "gender-swap them and see how different it would be" (Isaacson 2013). Like Grosser's extensions, Sucher extension is actively disrupting the social reality, and the effect can be disorienting. We have installed *JtP* at various times while teaching divergent, creative code classes and forgotten it was installed while using the browser. In some cases, the changes are obvious. In an article on Supreme Court Justice Ruth Bader Ginsberg, the statement "Public sightings of Ginsburg, who has his own action figure and nickname, the Notorious RBG, ripple across Twitter." The change the is immediately apparent. Other changes are easy to miss. Skimming the Wikipedia article for the history of Europe, this statement is jarring enough to cause a pause: "She was forced to withdraw. On the march back her army was harassed by Cossacks, and suffered disease and starvation. Only 20,000 of her women survived the campaign." To Sucher's point, this alteration of reality brings to the forefront that anyone trying to learn history is constantly, line by line, confronted by patriarchy. However, without remembering the extension is installed, there may be a delay in realizing the text has been altered. That delay could be described as the truth-default of reading web pages, referencing Truth-Default Theory. The dissonance between a major battle in history, and the idea that it was fought by women is enough to trigger a "Deception judgment" (Levine 2014), but in other cases it is possible to reach the end of a news article without realizing the extension has changed it significantly.

### *Sorry* Prototype

In developing our prototype ATD software, *Sorry*, the web extensions above provided some hint that the web interface can be changed without immediate notice, and offered approaches to constructing altered realities via fairly simple software. *Sorry* is a Firefox and Chrome extension that uses regular expressions in Javascript to find "I" statements and add the word "sorry" to them. Regular expressions are sequences of characters that match patterns (Goyvaerts 2017). As shown in Figure 1, the *Sorry* prototype finds phrases like "I am looking for…"; "I'm done with this"; and "I don't agree" and inserts the word "sorry." Those statements become "Sorry, I am looking for…"; "Sorry, I'm done with this"; and "Sorry, I don't agree."

```
var icontractionsstart = "^(I[ |'d|'ll|'m|'ve]+)\\b"; // matches I'X
statements at the beginning of string.
var icontractionsmiddle = "\\s(I[\\s|'d|'ll|'m|'ve]+)\\b"; // matches only "
I'm" in: If I am sad, I'm feeling furious.
var beginning = new RegExp(icontractionsstart, 'i');
var middle = new RegExp(icontractionsmiddle, 'gi');
```

Fig 1. Regular expressions in Javascript from Sorry prototype

The *Sorry* extension was originally developed as a creative code project. Creative coding is often described in terms of generating a visual aesthetic (Peppler and Kafai 2005). However, Mitchell and Brown describe it as "a discovery-based process consisting of exploration, iteration, and reflection, using code as a primary medium, towards a media artefact designed for an artistic context" (Mitchell and Bown 2013). *Sorry* is intended to make "I" statements seem apologetic, and craft an alternate reality for people who install the extension. By softening these statements, it serves as a counter to social media *hot takes*, or "piece[s] of deliberately provocative commentary [that are] based almost entirely on shallow moralizing" (Reeve 2015). In practice, it primarily changes the statement to seem apologetic, as if the author of the statement is self-doubting, or overly polite. Sometimes it seems sarcastic.

Unlike *Jailbreak the Patriarchy*, which targets all text on the web in order to foreground specific language, *Sorry* targets individual statements, and changes what the authors of those statements intended. This significant difference gave rise to the concept of using this approach to craft specific realities for individual users. Using regular expressions, it is theoretically possible to target only emails to and from specific people and to remove, change, or add specific affect. During the development of *Sorry* we frequently forgot that the extension was installed and found ourselves the victim of our own ATD attack, wondering why statements on social media seemed so apologetic and self-doubting.

## Conclusion

The concept of Ambient Tactical Deception was developed at the intersection of divergent creative code research and unfolding world events, particularly those involving cybersecurity and online social reality. A Washington Post article on Russian activity during the 2016 election describes the threat of information warfare, "Influence the information flow voters receive, and you'll eventually influence the government" (Klaas 2017). The article quotes Russian political scholar Igor Panarin, "influence can be achieved by information manipulation, disinformation, fabrication of information." What we have proposed in this paper is a potential new threat, a new approach to information warfare, and a new, potentially micro-targeted way to shape social reality. The approach is technologically feasible, and this paper has detailed multiple first steps toward crafting language in such a way as to alter a victim's social reality, particularly making a victim feel negatively about another person or other persons. A coordinated ATD attack could target  multiple people in an online political community or a business that employs online workers. In the field of cybersecurity, detailing possible exploitations of vulnerabilities in software (or "exploits") is intended to alert people about the risk, and present an opportunity to fix (or "patch") the vulnerability. In the case of ambient tactical deception, however, the vulnerabilities come from a complex intersection of human behavior and the alienation of online-only communication. A cybersecurity-based patch will, at best, involve tighter browser or device security. The patch for ambient social engineering, however, would require potential victims to distrust their software to a degree that does not seem possible, given how readily people grant permissions to apps on their phones, and their private information to social media companies. This paper, then, also serves as a general alert regarding the degree to which people are vulnerable when social discourse is limited to the degree required for contemporary, online-only communication.

## Discussion: Ambient Tactical Therapy

In an interview, Ben Grosser said that users of his *Facebook Demetricator* "talk about how the lack of numbers produce a calm, an ease; gives them a sense of relief, and makes Facebook seem less competitive" (Netburn 2012). *Jailbreak the Patriarchy* is intended to enlighten users and explore alternative, fictional gender realities. Even the *Sorry* prototype was originally intended to soften an increasingly polarized social media landscape. While this paper has focused on malicious use of automated text alteration, future research may consider whether there are other applications of this approach. Without the deception involved in hacking into a user's computer and installing malware, it may be possible to craft a more psychologically supportive environment for software users, and, as with FB Purity's text filer and Facebook's advertising preferences, remove distressing or triggering text and posts.

# References

Baldwin, T. 2018. *Ctrl Alt Delete: How Politics and the Media Crashed Our Democracy*. London, UK: Hurst. https://books.google.com/books?id=59ZjDwAAQBAJ.

Byron, Kristin. 2008. "Carrying Too Heavy a Load? The Communication and Miscommunication of Emotion by Email." *The Academy of Management Review* 33 (2). Academy of Management: 309–27. http://www.jstor.org/stable/20159399.

Danion, Jean-Marie, Françoise Kauffmann-Muller, Danielle Grangé, Marie-Agathe Zimmermann, and Philippe Greth. 1995. "Affective Valence of Words, Explicit and Implicit Memory in Clinical Depression." *Journal of Affective Disorders* 34 (3): 227–34. doi:https://doi.org/10.1016/0165-0327(95)00021-E.

Descartes, R, R Ariew, and D Cress. 2006. *Meditations, Objections, and Replies*. Hackett Classics. Indianapolis, IN: Hackett Publishing Company, Incorporated. https://books.google.com/books?id=vcpgDwAAQBAJ.

Dhamija, Rachna, J D Tygar, and Marti Hearst. 2006. "Why Phishing Works." In *Proceedings of the SIGCHI Conference on Human Factors in Computing Systems*, 581–90. CHI '06. New York, NY, USA: ACM. doi:10.1145/1124772.1124861.

Dimmick, John, Yan Chen, and Zhan Li. 2004. "Competition Between the Internet and Traditional News Media: The Gratification-Opportunities Niche Dimension." *Journal of Media Economics* 17 (1). Routledge: 19–33. doi:10.1207/s15327736me1701_2.

Experts Exchange. 2018. "Processing Power Compared." *Processing Power Compared*. https://pages.experts-exchange.com/processing-power-compared.

FB Purity. 2012. "Block Political / Sports Etc Posts with FB Purity's Custom Text Filter Word Lists | F.B. Purity – Cleans Up Facebook." *FB Purity*.

https://www.fbpurity.com/news/block-sports-political-posts-using-fb-puritys-custom-text-filter-word-lists/.

Felt, Adrienne Porter, Elizabeth Ha, Serge Egelman, Ariel Haney, Erika Chin, and David Wagner. 2012. "Android Permissions: User Attention, Comprehension, and Behavior." In *Proceedings of the Eighth Symposium on Usable Privacy and Security*, 3:1--3:14. SOUPS '12. New York, NY, USA: ACM. doi:10.1145/2335356.2335360.

Gendron, Maria, and Lisa Feldman Barrett. 2018. "Emotion Perception as Conceptual Synchrony." *Emotion Review* 10 (2): 101–10. doi:10.1177/1754073917705717.

Gordon, Whitson. 2010. "F. B. Purity Hides Annoying Facebook Applications and News Feed Updates." *LifeHacker*. https://lifehacker.com/5605377/f-b-purity-hides-annoying-facebook-applications-and-news-feed-updates.

Göritz, Anja S. 2007. "The Induction of Mood via the WWW." *Motivation and Emotion* 31 (1): 35–47. doi:10.1007/s11031-006-9047-4.

Goyvaerts, Jan. 2017. "Regular Expression Tutorial - Learn How to Use Regular Expressions." *Regular Expressions Tutorial*. https://www.regular-expressions.info/tutorial.html.

Grosser, Ben. 2018. "Facebook Demetricator | Benjamin Grosser." *About*. https://bengrosser.com/projects/facebook-demetricator/.

Hafner, Katie, and John Markoff. 1995. *Cyberpunk: Outlaws and Hackers on the Computer Frontier, Revised*. New York, NY: Simon and Schuster.

Haskins, Caroline. 2018. "Facebook Ad Micro-Targeting Can Manipulate Individual Politicians." *The Future*. https://theoutline.com/post/5411/facebook-ad-micro-targeting-can-manipulate-individual-politicians?zd=1&zi=y4kiv2mi.

Informate Mobile Intelligence. 2015. "International Smartphone Mobility Report – Dec. '14."

http://informatemi.com/blog/?p=75.

Isaacson, Betsy. 2013. "Jailbreak The Patriarchy Can Gender-Swap Everything You Read On
The Internet | HuffPost." *HuffPost*. https://www.huffingtonpost.com/2013/08/29/jailbreak-
the-patriarchy_n_3443654.html.

Klaas, Brian. 2017. "Stop Calling It 'Meddling.' It's Actually Information Warfare." *The
Washington Post*. https://www.washingtonpost.com/news/democracy-
post/wp/2018/07/17/stop-calling-it-meddling-its-actually-information-
warfare/?noredirect=on&utm_term=.a792bc4b8982.

Kopp, Carlo. 2003. "Shannon, Hypergames and Information Warfare." *Journal of Information
Warfare* 2 (2): 108–18.

Kozlowska, Hannah. 2018. "The Cambridge Analytica Scandal Affected Nearly 40 Million More
People than We Thought." *Quartz*.

Lam, C. 2011. "Linguistic Politeness in Student-Team Emails: Its Impact on Trust Between
Leaders and Members." *IEEE Transactions on Professional Communication* 54 (4): 360–
75. doi:10.1109/TPC.2011.2172669.

Levine, Timothy R. 2014. "Truth-Default Theory (TDT): A Theory of Human Deception and
Deception Detection." *Journal of Language and Social Psychology* 33 (4): 378–92.
doi:10.1177/0261927X14535916.

Lin, Herbert, and Jackie Kerr. 2017. "On Cyber-Enabled Information/Influence Warfare and
Manipulation On Cyber-Enabled Information/Influence Warfare and Manipulation." *Ssrn*,
1–29.

Mackiewicz, Jo, and Kathryn Riley. 2003. "The Technical Editor as Diplomat: Linguistic
Strategies for Balancing Clarity and Politeness." *Technical Communication* 50 (1).

Mitchell, Mark C, and Oliver Bown. 2013. "Towards a Creativity Support Tool in Processing: Understanding the Needs of Creative Coders." In *Proceedings of the 25th Australian Computer-Human Interaction Conference: Augmentation, Application, Innovation, Collaboration*, 143–46. OzCHI '13. New York, NY, USA: ACM. doi:10.1145/2541016.2541096.

Mozilla. 2018. "Confidentiality, Integrity, Availability." *Web Technologies for Developers*. https://developer.mozilla.org/en-US/docs/Web/Security/Information_Security_Basics/Confidentiality,_Integrity,_and_Availability.

Netburn, Deborah. 2012. "Facebook Demetricator May Be a Solution to Your 'likes' Addiction." *LA Times*. http://articles.latimes.com/2012/dec/05/business/la-fi-tn-facebook-demetricator-20121204.

Newman, Lily Hay. 2018. "No Title." *Wired*. https://www.wired.com/story/chrome-extension-malware/.

Opazo, B, D Whitteker, and C Shing. 2017. "Email Trouble: Secrets of Spoofing, the Dangers of Social Engineering, and How We Can Help." In *2017 13th International Conference on Natural Computation, Fuzzy Systems and Knowledge Discovery (ICNC-FSKD)*, 2812–17. doi:10.1109/FSKD.2017.8393226.

Peppler, K, and Y Kafai. 2005. "Creative Coding: Programming for Personal Expression." *Creative Coding*. doi:10.1.1.88.1191.

Perrin, Andrew, and Jingjing Jiang. 2018. "About a Quarter of U.S. Adults Say They Are 'Almost Constantly' Online." *FactTank*.

Reeve, Elspeth. 2015. "A History of the Hot Take." *The New Republic*.

https://newrepublic.com/article/121501/history-hot-take.

Rosenberg, Joyce. 2017. "Slack, Skype, Zoom: Remote Work the Norm Even at Small Firms."

 *AP Small Buinsess*. https://apnews.com/016fcf83396b4fe98c20356480269297.

Sadri, Fariba. 2011. "Ambient Intelligence: A Survey." *ACM Computuer Surveys* 43 (4). New

 York, NY, USA: ACM: 36:1--36:66. doi:10.1145/1978802.1978815.

Schatz, Daniel, Julie Wall, and Julie Wall. 2017. "Towards a More Representative Definition of

 Cyber Security Towards a More Representative Definition of Cyber Security." *The Journal*

 *of Digital Forensics, Security and Law (ADFSL)* 12 (2).

Shearer, Elisa, and Jeffrey Gottfried. 2017. "News Use Across Social Media Platforms 2017."

 *Journalism & Media*. http://www.journalism.org/2017/09/07/news-use-across-social-media-

 platforms-2017/.

Sloane, Garett. 2016. "Facebook Now Lets Users Block Ads That Stir Painful Memories | Digital

 - Ad Age." *AdAge*. https://adage.com/article/digital/facebook-lets-users-block-ads-stir-

 painful-memories/307193/.

Sucher, Danielle. 2015. "Jailbreak-the-Patriarchy." *Github*.

 https://github.com/DanielleSucher/Jailbreak-the-Patriarchy/blob/master/myscript.js.

The Conversation. 2014. "Awkward Pauses in Online Calls Make Us See People Differently."

 *Science+Technology*. https://theconversation.com/awkward-pauses-in-online-calls-make-

 us-see-people-differently-26073.

Verheyen, Christopher, and Anja S Göritz. 2009. "Plain Texts as an Online Mood-Induction

 Procedure." *Social Psychology* 40 (1): 6–15. doi:10.1027/1864-9335.40.1.6.

Vincent, James. 2018. "This Blessed Chrome Extension Replaces 'Elon Musk' with 'Grimes's

 Boyfriend'." *The Verge*. https://www.theverge.com/tldr/2018/5/10/17338984/elon-musk-

grimes-boyfriend-chrome-extension.

Williams, Emma J, and Danielle Polage. 2018. "How Persuasive Is Phishing Email? The Role of Authentic Design, Influence and Current Events in Email Judgements." *Behaviour & Information Technology* 0 (0). Taylor & Francis: 1–14. doi:10.1080/0144929X.2018.1519599.